\documentclass[showpacs,aps,prd,nofootinbib,floatfix,amsmath,amssymb]{revtex4}
\usepackage{graphicx}
\begin{document}

\title{Electric dipole and
magnetic quadrupole moments of the $W$ boson via a CP-violating
$HWW$ vertex in effective Lagrangians}
\author{J. Monta\~no}\affiliation{Departamento de
Matem\'aticas y F\'{\i}sica, Universidad Aut\'onoma de
Aguascalientes, Av. Universidad 940, C.P. 20100, Aguascalientes,
Ags., M\'exico}
\author{F. Ram\'{\i}rez-Zavaleta}
\affiliation{Departamento de
F\'{\i}sica, CINVESTAV, Apartado Postal 14-740, 07000, M\'exico, D.
F., M\'exico}
\author{G. Tavares-Velasco}
\author{J. J. Toscano}
\affiliation{Facultad de Ciencias F\'{\i}sico Matem\'aticas,
Benem\'erita Universidad Aut\'onoma de Puebla, Apartado Postal 1152,
Puebla, Pue., M\'exico}

\date{\today}

\begin{abstract}
The possibility of nonnegligible $W$ electric dipole
($\widetilde{\mu}_W$) and magnetic quadrupole ($\widetilde{Q}_W$)
moments induced by the most general $HWW$ vertex is examined via the
effective Lagrangian technique. It is assumed that new heavy
fermions induce an anomalous CP-odd component of the $HWW$ vertex,
which can be parametrized by an $SU_L(2)\times U_Y(1)$-invariant
dimension-six operator. This anomalous contribution, when combined
with the standard model CP-even contribution, lead to CP-odd
electromagnetic properties of the $W$ boson, which are characterized
by the form factors $\Delta \widetilde{\kappa}$ and $\Delta
\widetilde{Q}$. It is found that $\Delta \widetilde{\kappa}$ is
divergent, whereas $\Delta \widetilde{Q}$ is finite, which reflects
the fact that the latter cannot be generated at the one-loop level
in  any renormalizable theory. Assuming reasonable values for the
unknown parameters, we found that $\widetilde{\mu}_W\sim 3-6\times
10^{-21}$ e$\cdot$cm, which is eight orders of magnitude larger than
the SM prediction and close to the upper bound derived from the
neutron electric dipole moment. The estimated size of the somewhat
less-studied $\widetilde{Q}_W$ moment is of the order of $-10^{-36}$
e$\cdot$cm$^2$, which is fifteen orders of magnitude above the SM
contribution.
\end{abstract}


\maketitle

\section{Introduction}

It is well-known that a spin-$s$ particle associated with a no
self-conjugate field has $2s$ CP-odd permanent electromagnetic
moments. In particular, a charged \cite{Hagiwara} or neutral vector
particle \cite{Nieves,TT1}  has two electromagnetic moments, namely,
the electric dipole moment (EDM) and the magnetic quadrupole moment
(MQM). The scrutiny of these properties may provide relevant
information for our knowledge of CP violation, which still remains a
mysterious phenomenon whose experimental validity has been well
established via some flavor-changing processes such as the mixing of
the $K$ \cite{K} and $B$ \cite{B} hadrons. There is no yet
conclusive evidence that the origin of CP violation in
$K^0-\bar{K}^0$ mixing is  the Cabibbo-Kobayashi-Maskawa (CKM) phase
\cite{CKM}, which is the only source of CP violation in the
electroweak sector of the standard model (SM), but recent results
from $B$ factories at SLAC and KEK strongly suggest \cite{B,Soni}
that the dominant contribution to $B^0-\bar{B}^0$ mixing arises from
such a phase indeed. This means that, as far as $B$ hadron physics
is concerned, there is no much room left to detect new sources of CP
violation. On the other hand, diverse studies \cite{PospelovReview}
show that the CKM phase has a rather marginal impact on
flavor-diagonal processes such as the electric dipole moments of
elementary particles, which means that they could be highly
sensitive to new sources of CP violation. In fact, neither the
fermions  nor the $W$ gauge boson can have EDMs at the one-loop
level because a CP-violating phase cannot arise at this order since
the corresponding amplitudes depend only on the absolute value of
the CKM matrix elements. It has been shown \cite{SMPospelov1} that
the EDM of both quarks and $W$ boson vanishes also at the two-loop
level and appears first at the three-loop level
\cite{SMPospelov1,SMDarwing}. In the case of charged leptons,
without the presence of right-handed neutrinos the EDM is still more
suppressed and it can only be generated at the four-loop level or
higher orders \cite{SMPospelov12}. However, the existence of new
sources of CP violation in this sector is expected due to the recent
discovery of neutrinos masses and lepton mixing \cite{NM}. In
contrast, as far as the $W$ MQM is concerned, it has been shown
\cite{SMPospelov2} that it receives a tiny contribution at the
two-loop level in the SM.

Apart from the mere theoretical interest, the study of the EDM and
MQM of the $W$ boson is important from the experimental point of
view because they can induce large contributions to the EDM of the
fermions \cite{MQ}, such as the electron or the neutron, which may
be at the reach of low-energy experiments. It is thus worth
investigating new sources of CP violation beyond the SM. Although
these properties of the $W$ boson can only be generated through loop
effects within renormalizable theories, they may receive large
contributions in many SM extensions \cite{Marciano}. The only class
of models which can generate these quantities at the one-loop level
are those involving both left- and right-handed currents with
complex phases \cite{TT1}. In principle, this one-loop generated
effect would contribute dominantly to these $W$ properties, but it
could be strongly suppressed due to the presence of a tiny complex
phase, as occurs in left-right symmetric models(LRSM) \cite{LR} due
to experimental constraints on the $W_L-W_R$ mixing. Although the
presence of at least one fermionic loop involving a Dirac trace is
required to generate a term proportional to the Levi-Civita tensor
in the $WW\gamma$ vertex, the combination of fermion and scalar
fields may supply a potential source of CP violation in
flavor-diagonal processes, provided that their interactions involve
both scalar and pseudoscalar couplings. The simultaneous presence of
these types of Higgs-fermion couplings violate CP invariance, which
in turn can induce a trilinear Higgs-$W$ vertex with similar CP
property at the one-loop level. A $\phi WW$ vertex including a
linear combination of CP-even and CP-odd couplings is enough to
generate a CP-odd component in the on-shell vertex $WW\gamma$, which
would correspond to a two-loop effect in a renormalizable theory.
Although this source of CP violation is generated at the two-loop
level, it could give a contribution to the  CP-odd electromagnetic
moments of the $W$ boson larger than those induced by other
alternative sources. This mechanism does not depend crucially on the
existence of a complex phase since it is a direct consequence of the
presence of Higgs-fermion couplings that violate CP invariance in
the fundamental Lagrangian, which contrasts with the case of the CKM
phase or that arising from left- and right-handed charged currents.
Indeed, it is not necessary to go beyond the Fermi scale to
introduce the most general renormalizable CP-violating $\phi
\bar{f}f$ vertex. It arises for instance in the Yukawa sector of the
type-III two-Higgs doublet model (THDM) \cite{THDM-III}, where the
$\phi WW$ vertex is induced at the one-loop level as a linear
combination of CP-even and CP-odd couplings, which in turn generate
the most general on-shell $WW\gamma$ vertex including both CP-even
and  CP-odd dynamical structures at the two-loop level. Although
there are many types of Feynman diagrams contributing to the $W$
CP-odd properties in this model \cite{THDM}, the contribution from
the one-loop $\phi WW$ vertex generated by the Yukawa coupling
${\mathcal L}_Y=-\sum_i\phi \bar{\psi_i}({\mathit e}+i{\mathit
o}\gamma_5)\psi_i$ differs from any other source as it leads to a
finite and gauge invariant result by itself \cite{TT2}. The same
type of effect can arise from heavy fermions that may be present in
several SM extensions, and we will focus on this possibility.
Although SM extensions composed by very complex Higgs sectors are
common, it is worth emphasizing that most of them contain at least
one SM-like Higgs boson $H$. Such a Higgs boson is SM-like in the
sense that it is expected to be relatively light, with a mass of
order of the Fermi scale, and with tree level couplings presenting
small deviations from the SM that vanish in some appropriate limit.
In this work, we are interested in studying this class of
deviations, which may lead to the appearance of CP-odd
electromagnetic properties of the $W$ boson. More specifically, we
will concentrate on the most general $HWW$ vertex including both
CP-even and CP-odd components. Instead of focusing on a specific
model, it is convenient to parametrize this class of effects in a
model independent manner via the effective Lagrangian technique
\cite{EL}. It is assumed that these effects are induced by particles
that are much heavier than the Fermi scale and can thus be
integrated out in the generating functional. This framework is quite
appropriate to describe any physics phenomenon that is absent or
very suppressed in the SM. Apart from the advantages of working in a
model-independent fashion, this approach has some additional
advantages from the technical point of view. In particular, a
two-loop calculation, as the one we are interested in, can be
treated as a one-loop effect. Below, we will discuss the structure
of the effective $HWW$ vertex and its implications on the EDM and
MQM of the $W$ boson.

The rest of the paper is organized as follows. In Sec. II, the
$SU_L(2)\times U_Y(1)$-invariant effective Lagrangian description of
the $HWW$ vertex is presented and used to determine its impact on
the on-shell $WW\gamma$ vertex at the one-loop level within the
effective theory, whereas Secs. III and IV are devoted to discuss
our results and present our conclusions, respectively.

\section{The effective $HWW$ coupling and its contribution to
$WW\gamma$}A scheme that is well suited to analyze new physics
effects lying beyond the Fermi scale consists in introducing
$SU_L(2)\times U_Y(1)$-invariant  operators of dimension higher than
four that modify the SM dynamics. In particular, the structure of
the tree level $HWW$ vertex can be modified by introducing a
dimension-six effective Lagrangian given by
\begin{equation}
\label{el} {\mathcal L}_{eff}=gm_WHW^-_\mu
W^{+\mu}+\frac{\alpha_{HWW}}{\Lambda^2}(\Phi^\dag \Phi)W^i_{\mu
\nu}W^{i\mu \nu}+
\frac{\widetilde{\alpha}_{HWW}}{\widetilde{\Lambda}^2}(\Phi^\dag
\Phi)W^i_{\mu \nu}\widetilde{W}^{i\mu \nu},
\end{equation}
where $W^i_{\mu \nu}$ is the $SU(2)$ field strength,
$\widetilde{W}^i_{\mu \nu}=\frac{1}{2}\epsilon_{\mu \nu \alpha
\beta}W^{i\alpha \beta}$, and $\Phi$ is the standard model Higgs
doublet. The parameters $\alpha_{HWW}$ and
$\widetilde{\alpha}_{HWW}$ parametrize the details of the underlying
physics, and they could be determined once the fundamental theory is
known. On the other hand, $\Lambda$ and $\widetilde{\Lambda}$ are
new different physics scales, the latter being associated with
CP-violating effects. Such anomalous Higgs-$W$ interactions have
already been considered in the literature \cite{HWW,HWWNLC}. For
instance, they were introduced to analyze the CP structure of the
$HWW$ coupling at the CERN large hadron collider (LHC) \cite{HWW},
and more recently at next linear colliders (NLC) \cite{HWWNLC}.
Also, it is worth mentioning that this vertex already arises at the
one-loop level within the context of the SM provided that at least
one $W$ boson is off-shell \cite{SMHWW}. For the purpose of this
work, it is enough to concentrate on the first and last terms of the
Lagrangian (\ref{el}). Our main goal  is to study the impact of this
anomalous $HWW$ vertex on the CP-odd electromagnetic properties of
the $W$ boson, which are generated at the one-loop level in the
context of effective Lagrangians.

\begin{figure}
\centering
\includegraphics[width=1.5in]{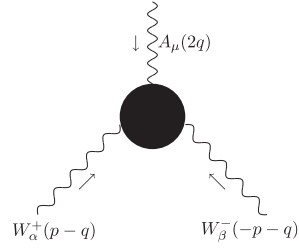}
\caption{\label{FIG1} The trilinear $WW\gamma$ vertex. The black
circle denotes anomalous contributions.}
\end{figure}

\begin{figure}
\centering
\includegraphics[width=2.5in]{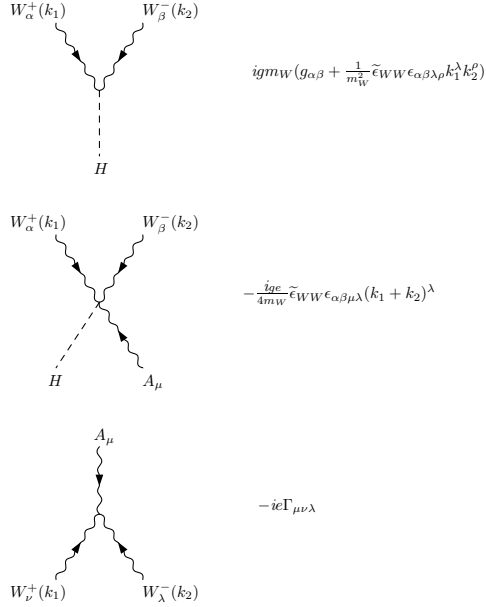}
\caption{\label{FIG2} Feynman rules for the vertices $HWW$,
$\gamma HWW$, and $WW\gamma$ in the unitary gauge.}
\end{figure}

\begin{figure}
\centering
\includegraphics[width=4in]{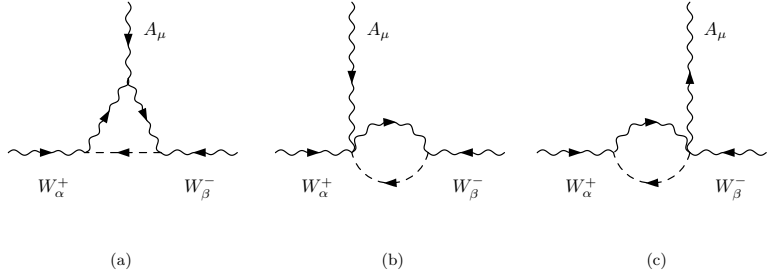}
\caption{\label{FIG3} Feynman diagrams contributing to the
$WW\gamma$ vertex in the unitary gauge.}
\end{figure}

The most general on-shell $WW\gamma$ vertex can be written as a
linear combination of CP-even and CP-odd electromagnetic gauge
structures \cite{Vertex}:
\begin{equation}
\Gamma_{\alpha \beta \mu}=ie\Big(\Gamma^{e}_{\alpha \beta
\mu}+\Gamma^{o}_{\alpha \beta \mu}\Big),
\end{equation}
where $\Gamma^{e}_{\alpha \beta \mu}$ ($\Gamma^{o}_{\alpha \beta
\mu}$ ) is the CP-even (CP-odd) component:
\begin{eqnarray}
\Gamma^{e}_{\alpha \beta \mu}&=&A\Big[2p_\mu g_{\alpha
\beta}+4(q_\beta g_{\alpha \mu}-q_\alpha g_{\beta \mu})\Big]
+2\Delta \kappa (q_\beta g_{\alpha \mu}-q_\alpha g_{\beta
\mu})+\frac{4\Delta Q}{m^2_W}p_\mu q_\alpha q_\beta,\\
\Gamma^o_{\alpha \beta \mu}&=&2\Delta
\widetilde{\kappa}\epsilon_{\alpha \beta \mu
\lambda}q^\lambda+\frac{4\Delta \widetilde{Q}}{m^2_W}q_\beta
\epsilon_{\alpha \mu \lambda \rho}p^\lambda q^\rho.
\end{eqnarray}
The notation and conventions used in these expressions are shown in
Fig. \ref{FIG1}. The $\Delta \kappa$ and $\Delta Q$ form factors
define the CP-conserving electromagnetic moments of the $W$ boson,
the magnetic dipole moment $\mu_W$ and the electric quadrupole
moment $Q_W$, through the following relations
\begin{eqnarray}
\mu_W&=&\frac{e}{2m_W}(2+\Delta \kappa), \\
Q_W&=&-\frac{e}{m^2_W}(1+\Delta \kappa+\Delta Q).
\end{eqnarray}
On the other hand, $\Delta \widetilde{\kappa}$ and $\Delta
\widetilde{Q}$ determine the CP-violating EDM and MQM:
\begin{eqnarray}
\widetilde{\mu}_W&=&\frac{e}{2m_W}\Delta \widetilde{\kappa}, \\
\widetilde{Q}_W&=&-\frac{e}{m^2_W}(\Delta
\widetilde{\kappa}+\Delta \widetilde{Q}).
\end{eqnarray}
We now turn to show that a $HWW$ vertex involving a linear
combination of both CP-even and CP-odd couplings induces the $\Delta
\widetilde{\kappa}$ and $\Delta \widetilde{Q}$ form factors at the
one-loop level. As already mentioned, this is a two-loop or higher
order effect since the effective operator $(\Phi^\dag \Phi)W^i_{\mu
\nu}\widetilde{W}^{i\mu \nu}$ can only be generated at one-loop or
higher orders by the underlying theory \cite{AEW}. Ignoring the term
associated with the coefficient $\alpha_{WW}$, the Higgs-$W$
interaction can be written, in the unitary gauge, as
\begin{equation}
{\mathcal L}_{HW}=gm_{W}HW^-_\mu W^{+\mu}+
\frac{g\widetilde{\epsilon}_{HWW}}{4m_W}H\Big[W^-_{\mu
\nu}\widetilde{W}^{+\mu \nu}+2ie(\widetilde{W}^+_{\mu \nu}A^\mu
W^{-\nu}-\widetilde{W}^-_{\mu \nu}A^\mu W^{+\nu})\Big],
\end{equation}
where $W^\pm_{\mu \nu}=\partial_\mu W^\pm_\nu-\partial_\nu
W^\pm_\mu$. We have introduced the definition
$\widetilde{\epsilon}_{HWW}=(v/\widetilde{\Lambda})^2\widetilde{\alpha}_{HWW}$,
being $v=246$ GeV the Fermi scale. Notice that, due to
$SU_L(2)\times U_Y(1)$-invariance, the $WW\gamma$ vertex receives
contributions from both the trilinear $HWW$ and quartic $HWW\gamma$
couplings. The Feynman rules needed to calculate the EDM and MQM are
shown in Fig. \ref{FIG2}, with the $\Gamma_{\mu \nu \lambda}$ tensor
given by
\begin{equation}
\Gamma_{\mu \nu \lambda}=-(k_1-k_2)_\mu g_{\nu
\lambda}-(k_1+2k_2)_\nu g_{\mu \lambda}+(2k_1+k_2)_\lambda g_{\mu
\nu}.
\end{equation}
In the unitary gauge, the CP-violating $HWW$ vertex contributes to
the $\Delta \widetilde{\kappa}$ and $\Delta \widetilde{Q}$ form
factors through the Feynman diagrams shown in Fig. \ref{FIG3}. It is
worth noting that the effective Lagrangian given in Eq. (\ref{el})
generates nonrenormalizable dimension-five $HWW$ and $\gamma HWW$
vertices, which yield a divergent contribution to $\Delta
\widetilde{\kappa}$. After evaluating the Feynman diagrams of  Fig.
\ref{FIG3}, we arrive at an ultraviolet divergent amplitude, which
is due to the presence of a nonrenormalizable interaction, thereby
requiring a  renormalization scheme. We have used the $\overline{\rm
MS}$ scheme with the renormalization scale
$\mu=\widetilde{\Lambda}$, which leads to a logarithmic dependence
of the form $\log(\widetilde{\Lambda}^2/m^2_W)$. After some algebra,
the CP-odd form factors can be written in terms of two-point
Passarino-Veltman scalar functions:
\begin{eqnarray}
\Delta
\widetilde{\kappa}&=&-\frac{\widetilde{\epsilon}_{HWW}\alpha}{4\pi
s^2_W}\frac{1}{48(x_H-4)}\Bigg[16(7+6(B_W-B_{WH}))
-8(23+21B_H+15B_W-36B_{WH})x_H\nonumber
\\
&&+3(25+26B_H+4B_W-30B_{WH})x^2_H-3(1+B_H-B_{WH})x^3_H
+9(x_H-4)x_H\log\left(\frac{\widetilde{\Lambda}^2}{m^2_W}\right)\Bigg],
\end{eqnarray}
\begin{equation}
\Delta \widetilde{Q}=-\frac{\widetilde{\epsilon}_{HWW}\alpha}{4\pi
s^2_W}\frac{1}{(x_H-4)}\Bigg[-4+(5+2(2B_H+B_W-3B_{WH}))x_H
-2(1+B_H-B_{WH})x^2_H\Bigg],
\end{equation}
where $x_H=(m_H/m_W)^2$, $B_H=B_0(0,m^2_H,m^2_H)$,
$B_W=B_0(0,m^2_W,m^2_W)$, and $B_{WH}=B_0(m^2_W,m^2_H,m^2_W)$.
Notice that the $\Delta \widetilde{Q}$ form factor is ultraviolet
finite, which is consistent with the fact that $\widetilde{\mu}_W$
and $\widetilde{Q}_W$ satisfy the relation $
2\widetilde{\mu}_W+m_W\widetilde{Q}_W=0$ in any renormalizable
theory as $\Delta \widetilde{Q}$ cannot arise at the one-loop level
\cite{LR,TT1}, but at higher orders. Since our calculation
represents a two-loop or higher order effect in the fundamental
theory, the contribution to $\Delta \widetilde{Q}$ must be
necessarily finite, in accordance with renormalization theory.

\section{Results and discussion.}  
\begin{figure}
\centering
\includegraphics[width=2.5in]{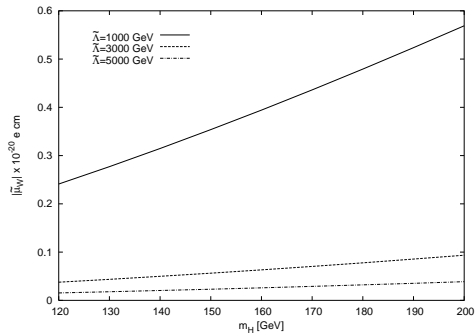}
\caption{\label{FIG4} The electric dipole moment $\widetilde{\mu}_W$
as a function of $m_H$ for $\widetilde{\Lambda}=1000$, 3000, and
5000 GeV.}
\end{figure}

\begin{figure}
\centering
\includegraphics[width=2.5in]{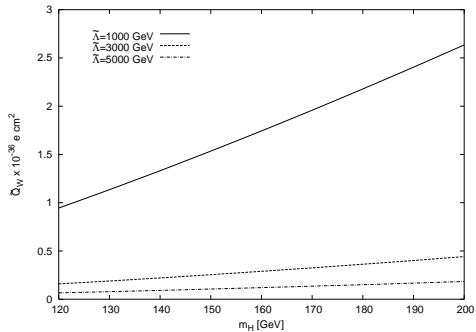}
\caption{\label{FIG5} The same as in Fig. \ref{FIG4} but now for the
magnetic quadrupole moment $\widetilde{Q}_W$.}
\end{figure}

\begin{figure}
\centering
\includegraphics[width=2.5in]{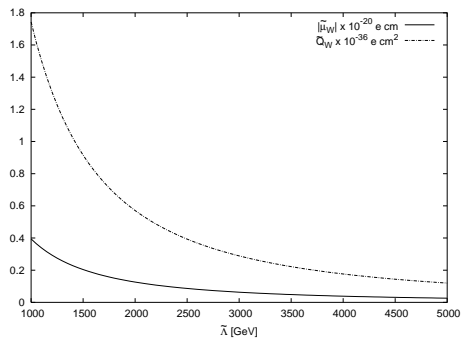}
\caption{\label{FIG6} The $\widetilde{\mu}_W$ and $\widetilde{Q}_W$
dependence on $\widetilde{\Lambda}$ for $m_H=160$ GeV.}
\end{figure}

We turn now to our numerical
results. The EDM and MQM of the $W$ boson depend on three free
parameters: the coupling constant $\widetilde{\alpha}_{HWW}$, the
new physics scale $\widetilde{\Lambda}$, and the Higgs boson mass
$m_H$. As already mentioned, the effective operator $(\Phi^\dag
\Phi)W^i_{\mu \nu}\widetilde{W}^{i\mu \nu}$ can only  be generated
at one-loop or higher orders by the fundamental theory \cite{AEW}.
Assuming that it is induced at the one-loop level, the
$\widetilde{\alpha}_{HWW}$ parameter must contain a factor of
$1/16\pi^2$ along with a $g$ coupling for each  gauge field. From
these considerations, it is reasonable to assume that
$\widetilde{\alpha}_{HWW}\sim g^2/(16\pi^2)f$, where
$f=f(v,\widetilde{\Lambda})$ is a dimensionless loop function, whose
specific structure depends on the details of the underlying physics.
Since the CP-violating effects are expected to be of decoupling
nature, $f$ is expected to be of the order $O(1)$ at most. In order
to make predictions, we will adopt a somewhat optimistic scenario,
which consists in assuming that $f\sim 1$. We will thus make
predictions under the assumption that $\widetilde{\epsilon}_{HWW}$
has the following form
\begin{equation}
\widetilde{\epsilon}_{HWW}=\Bigg(\frac{v}{\widetilde{\Lambda}}\Bigg)^2\frac{\alpha}{4\pi
s^2_W}.
\end{equation}
We now would like to analyze the behavior of $\widetilde{\mu}_W$ and
$\widetilde{Q}_W$  as a function of $m_H$ and $\widetilde{\Lambda}$.
The dependence of $\widetilde{\mu}_W$ ($\widetilde{Q}_W$) on the
Higgs boson mass is shown in Fig. \ref{FIG4} (\ref{FIG5}) for $m_H$
ranging between 120 GeV  and 200 GeV and for
$\widetilde{\Lambda}=$1, 3 and 5 TeV. From these Figures we can
observe  that $\widetilde{\mu}_W$ and $\widetilde{Q}_W$ are not very
sensitive to the Higgs boson mass. In fact,  for
$\widetilde{\Lambda}=1$ TeV, $\widetilde{\mu}_W$ ranges between
$0.25 \times 10^{-20}$ and $0.55 \times 10^{-20}$ e$\cdot$cm,
whereas $\widetilde{Q}_W$ goes from $1\times 10^{-36}$ to $2.5\times
10^{-36}$ e$\cdot$cm$^2$. As compared to the values obtained for
$\widetilde{\Lambda}=1$ TeV, $\widetilde{\mu}_W$ and
$\widetilde{Q}_W$ are decreased by a factor of $10^{-1}$ when
$\widetilde{\Lambda}=3$ TeV and  $5\times 10^{-1}$ when
$\widetilde{\Lambda}=5$ TeV. Finally, the behavior of the $W$
moments is shown in Fig. \ref{FIG6} as a function  of
$\widetilde{\Lambda}$  and for $m_H=160$ GeV.

It is worth comparing our results with those obtained in other
scenarios. To begin with, we would like to discuss the SM
predictions for $\widetilde{\mu}_W$ and $\widetilde{Q}_W$. As
already noted, the lowest order nonzero contribution to
$\widetilde{\mu}_W$ arises at the three-loop level, whereas
$\widetilde{Q}_W$ appears up to the two-loop order. At the lowest
order, $\widetilde{\mu}_W$ has been estimated to be smaller than
about $10^{-29}$ e$\cdot$cm \cite{SMDarwing,MoreSMEDM}. As far as
$\widetilde{Q}_W$ is concerned, it has been estimated to be about
$-10^{-51}$ e$\cdot$cm$^2$ \cite{SMPospelov2}. In contrast, some SM
extensions predict values for $\widetilde{\mu}_W$ that are several
orders of magnitude larger than the SM one. It must be emphasized
here that all these studies have focused only on
$\widetilde{\mu}_W$. For instance, a value of $10^{-22}$ e$\cdot$cm
was estimated for $\widetilde{\mu}_W$ in LRSM \cite{SMDarwing,LR},
and similar results were found in supersymmetric models, which
induce this moment via one-loop diagrams mediated by charginos and
neutralinos \cite{SMDarwing,SUSY}. Also,  a nonzero
$\widetilde{\mu}_W$ can arise through two-loop graphs in multi-Higgs
models \cite{Weinberg}. Explicit calculations carried out within the
context of THDMs show that $\widetilde{\mu}_W\sim 10^{-20}-10^{-21}$
e$\cdot$cm \cite{THDM}. A similar value was found in the context of
the so-called $331$ models, which induce a nonzero
$\widetilde{\mu}_W$ via two-loop graphs similar to the ones of THDM
\cite{331}. From these results, we can conclude that our
model-independent estimation for $\widetilde{\mu}_W$ lies within the
range of the predictions obtained from most popular renormalizable
theories.

It is well known that the $W$ EDM and MQM can induce important
contributions to the EDM of light fermions. This fact was exploited
by the authors of Ref. \cite{MQ}, who used the experimental upper
bound on the neutron EDM, $d_n<10^{-25}$ e$\cdot$cm, to obtain the
upper bound $\widetilde{\mu}_W<10^{-20}$ e$\cdot$cm. Our prediction
for $\widetilde{\mu}_W$, which is eight orders of magnitude larger
than that of the SM, is consistent with this upper bound. As far as
$\widetilde{Q}_W$ is concerned, currently there is no indirect
experimental upper bound, but we would like to emphasize that our
result is quite remarkable as predicts that new physics effects are
capable of enhancing $\widetilde{Q}_W$  up to fifteen orders of
magnitude above the SM contribution.

As far as the direct observation of the CP-odd structure of the
$WW\gamma$ vertex is concerned, the prospect of the NLC and CLIC
\cite{NLC} have triggered the interest in the $e^+e^-\to W^-W^+$
reaction as an efficient tool to produce large quantities of $W$
boson pairs, which would allow one to study new physics effects on
the $WWV$ ($V=\gamma, Z$) vertex. The possibility of extracting
CP-odd asymmetries from these colliders has been examined by several
authors, mainly in a model-independent approach via effective
Lagrangians \cite{ELSC1}. These studies suggest that an effect in
$\widetilde{\mu}_W$ at the level of $10^{-20}$ e$\cdot$cm may be at
the experimental reach. Also, it is expected that careful studies on
the polar and azimuthal distributions of lepton-antilepton pairs
produced in $W$ decays may further enhance the constraints on the
size of $\widetilde{\mu}_W$ \cite{ELSC2}.

\section{Conclusions}
The origin of CP violation is a fascinating open problem worthwhile
of both theoretical and experimental attention. Although
flavor-diagonal CP-violating processes, such as the electric dipole
and magnetic quadrupole moments of the $W$ boson, are not induced by
the CKM mechanism, their presence cannot be dismissed since they may
represent a genuine effect of physics lying beyond the Fermi scale.
This is suggested by many SM extensions predicting the existence of
new sources of CP violation that may give sizeable contributions to
the static quantities of the $W$ boson. As far as renormalizable
theories are concerned, the Yukawa sector of several SM extensions
seems to be a potential source of CP violation since it may involve
the presence of Higgs bosons with both scalar and pseudoscalar
couplings to the fermions. In this work, we examined such a
possibility via the model-independent approach of effective
Lagrangians. We assumed the existence of new heavy fermions that
generate the most general $HWW$ vertex at the one-loop level. Such a
CP-violating nonrenormalizable interaction was parametrized through
an $SU_L(2)\times U_Y(1)$-invariant operator and combined with the
usual CP-even SM coupling to calculate the one-loop contribution to
the CP-odd structure of the on-shell $WW\gamma$ vertex. The fact
that the dimension-six operator can only be generated at the
one-loop level means that the CP-odd property of the $WW\gamma$
vertex is a two-loop effect in the fundamental theory. Explicit
expressions for the form factors $\Delta \widetilde{\kappa}$ and
$\Delta \widetilde{Q}$, which define the EDM and MQM of the $W$
boson, were derived. An appropriate renormalization scheme was
adopted in order to renormalize the divergent form factor $\Delta
\widetilde{\kappa}$. As for the form factor $\Delta \widetilde{Q}$,
its nondivergent nature reflects the fact that it cannot arise at
the one-loop level in any renormalizable theory. Assuming reasonable
values for the unknown parameters of the effective theory, the
estimated values of the CP-odd electromagnetic moments are
$\widetilde{\mu}_W \sim 3-6\times 10^{-20}$ e$\cdot$cm and
$\widetilde{Q}_W\sim -10^{-36}$ e$\cdot$cm$^2$. These values are
eight and fifteen orders of magnitude above the respective SM
contributions. To our knowledge, this model-independent estimation
for the size of the $\widetilde{Q}_W$ is the first one obtained in
theories beyond the SM. On the other hand, the value predicted for
$\widetilde{\mu}_W$ is of the same order of magnitude or larger than
those predicted by some SM extensions, and it is consistent with the
existing indirect upper bound derived from the neutron electric
dipole moment.

\acknowledgments{We acknowledge  support from CONACYT under grant
U44515-F and SNI (M\' exico).}

\end{document}